\documentclass[letter]{aa}
\usepackage{graphicx}
\usepackage{multicol}
\usepackage{txfonts}
\usepackage{hyperref}

\newcommand{\omo}{\Omega_{m,0}}
\newcommand{\olo}{\Omega_{\Lambda,0}}
\newcommand{\ota}{\Omega_{ta}}

\newcommand{\rco}{\rho_{c,0}}

\begin{document}


\title{Turnaround density as a probe of the cosmological constant}

   \author{Vasiliki Pavlidou \inst{1}\fnmsep\inst{2}\fnmsep\thanks{E-mail: pavlidou@physics.uoc.gr},
          Giorgos Korkidis,
          \inst{1}\fnmsep\inst{2},
                    Theodore N. Tomaras\inst{1},
          Dimitrios Tanoglidis\inst{1}\fnmsep\inst{3}
          }
    
    \authorrunning{V. Pavlidou et al.}

   \institute{Department of Physics and Institute for Theoretical and
     Computational Physics, University of Crete,  GR-70013, Heraklio, Greece
         \and
             Institute of Astrophysics, Foundation for Research and Technology – Hellas, Vassilika Vouton, GR-70013 Heraklio, Greece
             \and 
             Department of Astronomy and Astrophysics and KICP, University of Chicago, Chicago, IL 60637, USA
             }

\date{}
\abstract
{  
Spherical collapse predicts that a single value of the turnaround density (average
matter density within the scale on which a structure
detaches from the Hubble flow) characterizes all cosmic structures at
the same redshift. It has been recently shown by Korkidis et al. that this feature
persists in complex non-spherical galaxy clusters identified in N-body simulations. 
 Here we show that the low-redshift evolution of the turnaround density constrains the cosmological  parameters, and that it can be used to derive a local constraint on $\olo$ alone, independent of $\omo$.
The turnaround density thus provides a promising new way to
exploit upcoming large cosmological datasets. }

\maketitle

\section{Introduction}
 Despite the numerous successes of concordance $\Lambda$CDM cosmology,
increasingly accurate cosmological datasets are
 starting to reveal tensions  (see
 e.g. \citealp{cepheids1,cepheids2, Lensing, LensingExtensions, Hu,
   PlanckCosmology2018, Charnock, Hu2, AH19, Riess, DDE, divalentino2019}). 
   Additionally, our
 evidence for the existence of vacuum energy, whether in the form of
 a cosmological constant $\Lambda$ or not, remains indirect, with different
 datasets constraining primarily the relation between the present-day
 values of the matter and $\Lambda$ density parameters ($\omo$ and
 $\olo$, respectively), rather than $\olo$
 alone. This is in contrast to $\omo$, to which certain datasets (e.g.
cluster abundances, baryon acoustic oscillations) are almost
 exclusively sensitive, independently of the value or existence of
 $\olo$. This periodically leads to a critical revisitation of the
 strength of the evidence that $\olo \neq 0$ \citep{Rocky2006, BR2012,
   sarkar2016, dam2017,CMRS2019,Kang2020}, also fuelled by the lack of generally-accepted fundamental-physics--driven candidates for the nature of vacuum energy. In this context, yet-unexplored probes of the cosmological
parameters can provide new insights to the cosmological model.

In recent years, considerable 
attention has been given to the properties of cosmic structures on the
largest scales as a means to locally probe cosmology and alternative
theories of gravity (e.g., \citealp{cuesta1, cuesta2, TC2008, PT14,
  PTT2014, splashback, Dragan,boundary1, Susmita2018, outskirts_review}). The turnaround radius (the scale on which a cosmic structure detaches from the Hubble flow) has been the focus of many such studies \citep{PTT2014, TPT2015, TPT2016, LeeLi2017,
  Sourav2017,Faraoni2018, Dia2019, Santa2019, Lopes2019, Wong19}.
The turnaround radius can be measured kinematically in any galaxy
cluster, as the boundary between the cluster and the expanding
Universe. Spherical collapse predicts that all structures turning
around at some cosmic epoch share a characteristic average density
within the turnaround radius, the {\em
  turnaround density} $\rho_{ta}$. \citet{Ketal2019} have shown, using
N-body simulations, that a
single turnaround radius also meaningfully describes simulated galaxy clusters
with realistic shapes, and that the average matter density within
that turnaround radius has a narrow distribution around a
characteristic value for clusters of all masses, 
consistent
with the predictions of spherical collapse. 

The turnaround density is sensitive to the presence of a cosmological constant
$\Lambda$. Once the effect of $\Lambda$ becomes dominant over the
gravitational self-attraction of matter, it halts structure growth
(\citealp{Busha1,Busha2,PT14,TPT2015}).  As a consequence, in an ever-expanding
Universe with $\Lambda$, not all overdensities are destined to
eventually 
detach from the Hubble flow (e.g. \citealp{PF05}),  and $\rho_{ta}$ has a hard lower bound of
$2\rho_\Lambda = 2(\Lambda c^2/8\pi G)$ \citep{PT14}.
The evolution of $\rho_{ta}$ thus changes between
early and late cosmic times in a cosmology-revealing manner. 
In $\Lambda$CDM, at early times, when matter dominates, $\rho_{ta}$ falls as $ a^{-3}$, where $a=(1+z)^{-1}$ is the
scale factor of the Universe  (see e.g. \citealp{Padma,Peebles}; and \S\ref{evolution}). At late times, when $\Lambda$
dominates, $\rho_{ta} \rightarrow 2\rho_\Lambda \propto
a^0$ (Fig.~\ref{rho-a}, black solid line). In contrast, in an open matter-only Universe,
$\rho_{ta}$ decreases without bound, as $a^{-3}$ while matter
dominates, and as $a^{-2}$ when curvature takes over
(Fig.~\ref{rho-a}, red dashed line).  For the present cosmic epoch,
concordance $\Lambda$CDM predicts that $\rho_{ta}\propto a^{-1.5}$ (Fig.~\ref{rho-a}, dotted black line),
already shallower than the asymptotic late-time behavior of a
matter-only Universe. It is
therefore reasonable to expect that a measurement of  the evolution
of  $\rho_{ta}$ with redshift could provide evidence for the existence
of $\Lambda$. This result is independent of the (universal for all
cosmologies) early-times behavior of $\rho_{ta}$, so observations at
low redshifts would be sufficient to establish it. The turnaround density could thus provide a “local” probe of the cosmological parameters, demonstrating the existence of dark energy by using its effect on scales much smaller than the observable universe, and structures located at low redshifts.

In this letter, we explore the type of constraints that could be
placed on the cosmological parameters $\omo$ and $\olo$ by 
measurements of the present-day value of the turnaround density,
$\rho_{ta,0}$, and of its present-day rate of change with redshift
$\left. d\rho_{ta}/dz \right|_0$. For the predictions of the
concordance $\Lambda$CDM model we use the 2018 Planck cosmological
parameters \citep{PlanckCosmology2018}: $\omo=0.315$, $\olo=1-\omo$,
$H_0=67.4 {\rm \, km/s/Mpc.}$ This choice has no qualitative effect on our conclusions.

\section{Evolution of $\rho_{ta}$}\label{evolution}
We consider a spherical shell of evolving radius $R_s$ destined to
eventually turn around, surrounding a single
spherical perturbation in an otherwise homogeneous and isotropic
Universe. We consider the background Universe on a scale $R$ large enough
so that the perturbation alters negligibly its expansion properties.

The evolution of both the background Universe and the shell will be
described by a Friedmann equation, each with a different curvature
constant $\kappa$. We consider shells that turn around late enough so
that their turnaround radius is much larger than their size at
matter-radiation equality, and we therefore only consider the matter,
curvature, and cosmological constant contributions to the Friedmann
equation. For the background Universe we write:  
\begin{equation}\label{Friedmann_Universe}
\left(\frac{\dot{R}}{R}\right)^2 = 
\frac{8\pi G}{3} \rho_m + \frac{\Lambda c^2}{3} 
- \frac{\kappa_U c^2}{R^2}\,.
\end{equation}
From Eq.~\ref{Friedmann_Universe} we can obtain $\kappa_U$ in terms of the present-day value of the Hubble
parameter, 
$H_0=\dot{R}|_0/R_0$(quantities with subscript $0$ refer to $z=0$): \begin{equation}
\kappa_Uc^2 = - R_0^2 \left[H_0^2 - \frac{8\pi G}{3} \rho_{m,0}
- \frac{\Lambda c^2}{3} \right]\,.
\end{equation}
We define the present-day critical density $\rco=3H_0^2/8\pi G$, and
the background-Universe scale factor $a=R/R_0=1/(1+z)$ (so that $a=1$
today and $\rho_{m}=\rho_{m,0}a^{-3}$). Then, substituting in (\ref{Friedmann_Universe}) yields the Friedmann equation in its most frequently encountered form, 
\begin{equation}\label{usual}
\left(\frac{\dot{a}}{a}\right)^2 = 
H_0^2 \left[ \omo a^{-3} + \olo +(1-\omo-\olo)a^{-2}\right]
\end{equation}
where $\olo = \Lambda c^2/3H_0^2$. For the spherical shell around the perturbation we write: 
\begin{equation}\label{Friedmann_Perturbation}
\left(\frac{\dot{R_s}}{R_s}\right)^2 = 
\frac{8\pi G}{3} \rho_{m,s} + \frac{\Lambda c^2}{3} 
- \frac{\kappa_s c^2}{R_s^2}\,.
\end{equation}
\begin{figure}[tb!]
    \centering
    \includegraphics[width=0.99\columnwidth,clip]{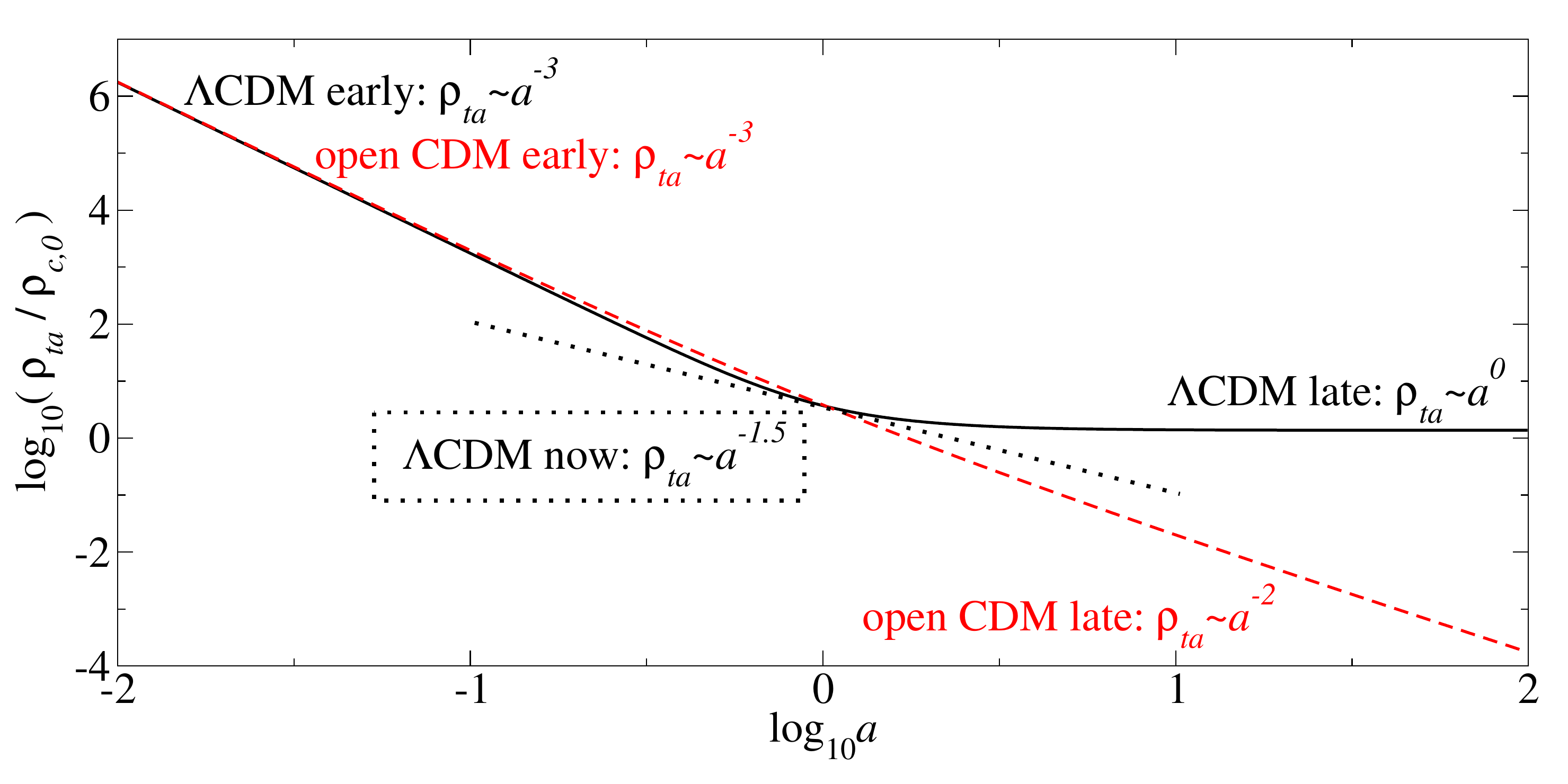}
    \includegraphics[width=0.99\columnwidth,clip]{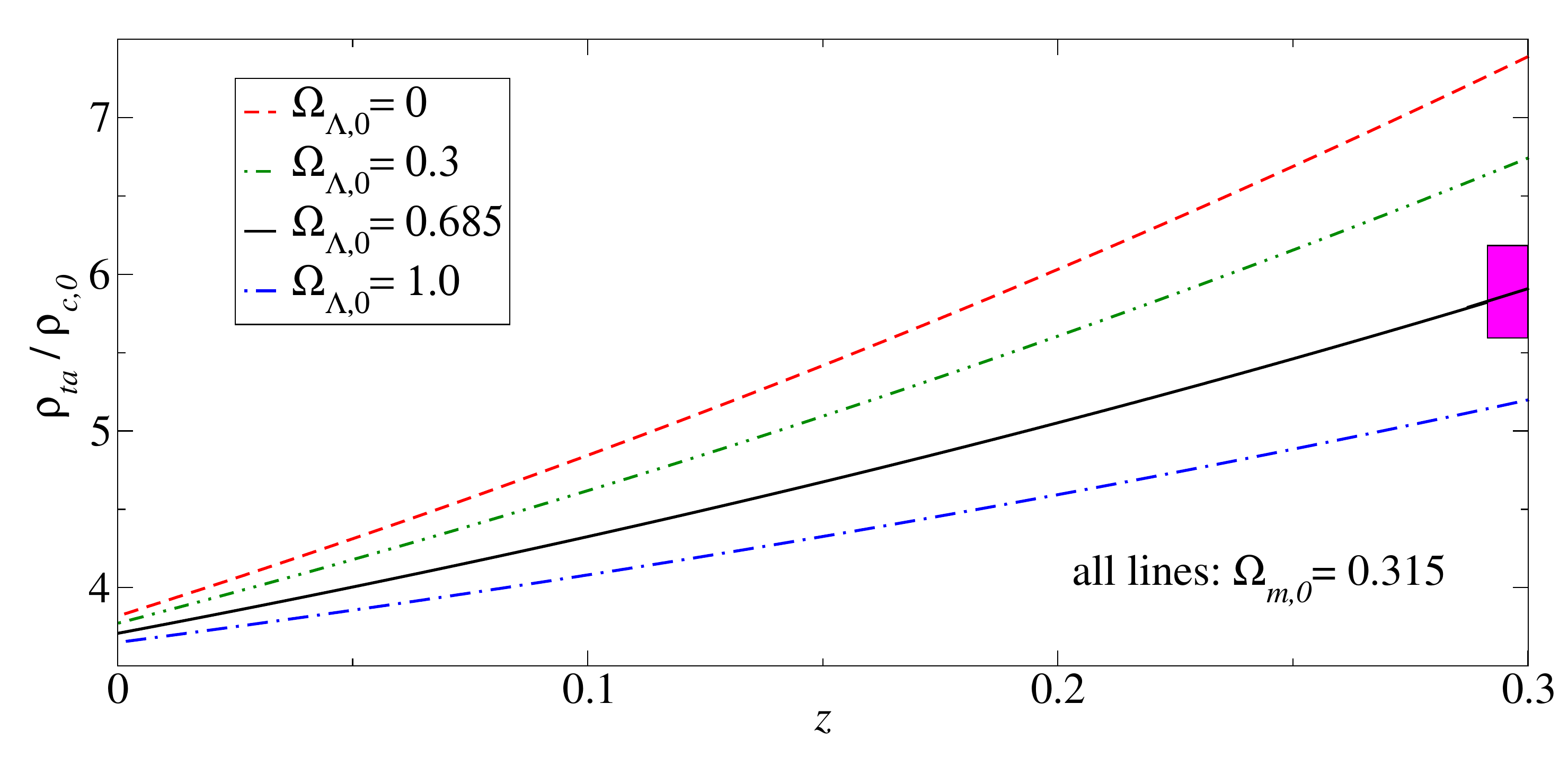}

    \caption{Upper panel: evolution with scale factor $a$ of $\rho_{ta}$ in units of
  the present-day critical density, $\rho_{\rm c,0}$. Black solid
  line: flat  $\Lambda$CDM cosmology with $\omo=0.315$
  \citep{PlanckCosmology2018}. Red dashed line:
  matter-only open cosmology, with $\omo=0.315$, $\olo=0$. Black dotted
  line: present-day ($a=1$) tangent to the black solid line, with
  slope $\rho_{ta}\sim a^{-1.5}$, shallower than the asymptotic
  behavior of an $\olo=0$ Universe.  Lower panel:
  evolution of $\rho_{ta}$ with redshift $z$, for $\omo = 0.315$ and
  different values of $\olo$. The magenta shaded box shows the accuracy that
  can be achieved by measuring $\rho_{ta}$ in 100 clusters at $z=0.3$ with
 fractional uncertainty of 50\% in each, and is indicative of
  the discriminating power of such an experiment. }
\label{rho-a}
\end{figure}
In this case, we obtain $\kappa_s$ by considering the state of the
shell at its time of turnaround, $a_{ta}$\footnote{If a structure is
  observed at redshift $z$, $a_{ta}=(1+z)^{-1}$; a shell achieving
  turnaround at that time can always be identified, as the scale on
  which the boundary of the structure joins the Hubble flow.}. Then,
$\dot{R_s}=0$; the size of the shell is equal to the turnaround radius, $R_{s,ta}$; and
its enclosed mass density is equal to the turnaround density $\rho_{ta}$ at  $a_{ta}$:
\begin{equation} 
\kappa_s c^2 = R_{s,ta}^2 \left(\frac{8\pi G}{3}\rho_{ta} + \frac{\Lambda c^2}{3}\right)\,.
\end{equation}
Substituting in (\ref{Friedmann_Perturbation}), defining the shell scale factor $a_s = R_s/R_{s,ta}$ (so that $a_s=1$ at the time of turnaround and $\rho_{m,s}=\rho_{ta}a_s^{-3}$), and measuring densities in units of the background-Universe critical density $\rco$ yields:
\begin{equation}\label{unusual}
\left(\frac{\dot{a_s}}{a_s}\right)^2 = 
H_0^2 \left[ \ota a_s^{-3} + \olo -(\ota+\olo)a_s^{-2}\right]\,,
\end{equation}
where $\ota=\rho_{ta}/\rho_{c,0}$. $\ota$ is a function $a_{ta}$, which in turn depends
on the initial overdensity within the shell: initially denser perturbations turn around
earlier. 
Dividing Eq.~(\ref{unusual}) by Eq.~(\ref{usual}) and taking the
positive square root (since for $a_s\leq 1$ both Universe and perturbation expand) we obtain
\begin{equation}
    \frac{da_s}{da} = \frac{a_s}{a}
    \sqrt{
    \frac{\ota a_s^{-3} + \olo -(\ota+\olo)a_s^{-2}}
    {\omo a^{-3} + \olo +(1-\omo-\olo)a^{-2}}
    }\,.
\end{equation}

The turnaround density $\ota$ as a function of turnaround time $a_{ta}$ can be obtained by integration of the perturbation scale factor $a_s$ from $0$ to $1$ and the Universe scale factor $a$ from $0$ to $a_{ta}$:
\begin{eqnarray}
&&\int_0^1    \frac{d a_s}{\sqrt{\ota (a_s^{-1}-1) + \olo(a_s^2-1)}} =\nonumber \\
&& \,\,\,\,
=\int_0^{a_{ta}} \frac{da}{\sqrt{\omo a^{-1} + \olo a^2 +(1-\omo -\olo)}}\,.
\label{functionf}
\end{eqnarray}
The result of this integration is plotted in the upper panel of
Fig.~\ref{rho-a}, for a flat $\Lambda$CDM cosmology with $\omo=0.315$
(black solid line), and an open CDM universe with $\omo=0.315$ and
$\olo=0$ (red dashed line). The present-day slope of the scaling is
shown with the dotted line. Remarkably, the present cosmic epoch
coincides with the era of transition between asymptotic behaviors. For
this reason, $\ota(z)$ curves for different values of $\olo$ 
deviate from each other very quickly, already at low $z$, as shown in
the lower panel of Fig.~\ref{rho-a}. 

In principle, two exact measurements of $\ota$ at two different
redshifts would uniquely determine $\omo$ and $\olo$. 
Alternatively, if $\ota$ is measured at
some redshift (e.g.  $z=0$, $a_{ta}=1$), then  Eq.~(\ref{functionf})
yields a constraint on the relative values of  $\omo$ and $\olo$.
A measurement of the present-day value of $\ota$
is most sensitive to the value of $\omo$
[see Fig.~\ref{newline}, red contours; note also in the lower panel of Fig.~\ref{rho-a} that different $\olo$
yield very similar present-day $\ota$ when $\omo$ remains the same]. 

The lower panel of Fig.~\ref{newline} shows that the present-day value
of the rate of change of
$\ota$ with $z$ for a given value of $\omo$ is very sensitive to
$\olo$. This implies that 
$\left. d\ota/dz\right|_0 =-\left. d\ota/da_{ta}\right|_0 $ would be a useful
cosmological observable. 
We can obtain a prediction for $d\ota/da_{ta}$ by differentiating Eq.~(\ref{functionf}) with respect to $a_{ta}$:
\begin{eqnarray}
&&-\frac{1}{2}\frac{d\ota}{da_{ta}}\int_0^1    \frac{d a_s (a_s^{-1}-1)}
{\left[\ota (a_s^{-1}-1) + \olo(a_s^2-1)\right]^{3/2}} =\nonumber \\
&& \,\,\,\,\,\,\,\,\,\,\,\,
=\frac{1}{\sqrt{\omo a_{ta}^{-1} + \olo a_{ta}^2 +(1-\omo -\olo)}}\,.
\label{functiong}
\end{eqnarray}
For the present cosmic epoch ($a_{ta}=1$) Eq.~(\ref{functiong}) becomes
\begin{equation}\label{magick}
-\frac{1}{2}\left.\frac{d\ota}{da_{ta}}\right|_0
\int_0^1    \frac{d a_s (a_s^{-1}-1)}
{\left[{\ota}_0 (a_s^{-1}-1) + \olo(a_s^2-1)\right]^{3/2}}
=1\,,
\end{equation}
independent of $\omo$.

If a measurement is then made of $\ota$ in a sufficient number of
clusters at a range of (low) redshifts, then both the slope  
$\left. d\ota/dz\right|_0 $ and
the intercept ${\ota}_0$ of the scaling can be derived. A constraint on $\olo$ alone can thus be obtained from Eq.~(\ref{magick}) [see
Fig.~\ref{newline}, purple contours].

\section{Possible constraints on cosmological
  parameters}\label{constraints}

Here we present a simplified estimate of the potential accuracy of a
measurement of ${\ota}_0$ and $\left. d\ota/da_{ta}\right|_0 $, and of
the associated inference of the cosmological parameters $\omo$ and
$\olo$, under the following three assumptions.
1. Errors are dominated by statistical uncertainties. 2. 
$\ota(z)$ behaves approximately linearly with $z$ in the low
redshift range we will consider. 3.  The redshift
 distribution of measured clusters follows\footnote{This is equivalent
   to assuming that distances $\propto z$ and a constant
 number density of clusters in the nearby 
Universe. Such a distribution approximates well the redshift
distribution of clusters in e.g.
\citet{WHLClusters} out to $z\sim 0.2$. For higher redshifts the
number of clusters in \citet{WHLClusters} grows more slowly with $z$, which  for a
fixed number of clusters results in
tighter constraints.} $dn/dz\propto z^2$.
\begin{figure}[tb!]
    \centering
    \includegraphics[width=0.9\columnwidth,clip]{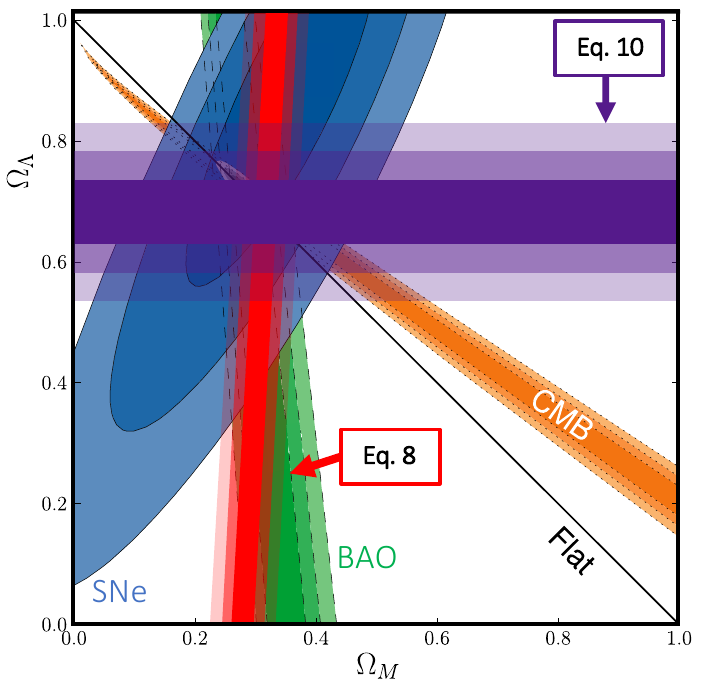}
\caption{Constraints on $\omo$ and $\olo$ from the cosmic microwave
  background (CMB, orange
  contours, WMAP, \citealp{cmbdata}),
  supernovae (blue contours, Union 2 SN Ia compilation,
  \citealp{ConstraintsPlot}), and baryon acoustic oscillations (green
  contours, SDSS, \citealp{BAOdata}); from \citealp{ConstraintsPlot},
  Fig.~10). The red and purple contours correspond to projected $1$, $2$, and $3\sigma$
  constraints implied by Eq.~(\ref{functionf}) and
  Eq.~(\ref{magick}), respectively, using a presumed  high-accuracy measurement of the
  evolution of $\ota$ at the low-redshift Universe ($\sim 42,000$
  galaxy clusters at $z\leq 0.3$, with an individual-cluster $\ota$
  uncertainty of $50\%$, see \S \ref{constraints}), yielding a
  1.5\% accuracy estimate  of ${\ota}_0$ and a   $3.5\%$ estimate of  $d\ota/dz|_0$. 
  We have assumed that $\ota$ evolves with $z$ as predicted by flat $\omo=0.315$, $\Lambda$CDM cosmology.  At this level of accuracy, $\olo>0$ could be established at a 14$\sigma$ confidence level from the turnaround density data alone.}
\label{newline}
\end{figure}

We can obtain an estimate of the
uncertainties in $d\ota/dz|_0 $ and ${\ota}_0$ 
by considering the errors in the slope and intercept
of a linear regression fit to $\ota (z) \approx  {\ota}_0+(d\ota/dz|_0)z$ of a
sample of $n$ measurements of $(z_i, {\ota}_i)$ in individual galaxy
clusters with $z\leq z_{max}$. In each cluster, ${\ota}_i$ is measured
with some uncertainty $\sigma_{{\ota},i}$. In linear regression, the
standard error of the slope, $\sigma_{d\ota/dz|_0}$, is $\sqrt{\left(\sum_{i=1}^n \epsilon_i^2/n\right)/\left[(n-2){\rm
      Var}(z)\right]}$, where $\epsilon_i$ are the regression residuals and ${\rm
Var}(z)$ the variance of the independent variable. $\sum_{i=1}^n
\epsilon_i^2/n$ is equal to $\langle \sigma^2_{{\ota},i}\rangle$. If  $\ota$ in all clusters can be measured with
the same fractional accuracy $f$, then $\sum_{i=1}^n \epsilon_i^2/n =
\langle f^2\ota^2(z)\rangle= f^2 \langle \left[
  {\ota}_0+(d\ota/dz|_0)z\right]^2 \rangle $.
To calculate both the latter average and ${\rm Var}(z)$, we use 
 $dn/dz\sim z^2$.  Finally, dividing by $d\ota/dz|_0$,  we obtain the fractional uncertainty 
of the slope:
\begin{eqnarray}
  \frac{\sigma_{d\ota/dz|_0}}{d\ota/dz|_0 }
&\sim&
\frac{5f}{\sqrt{n-2}} \sqrt{
\frac{{\ota^2}_0/z^2_{max}}{(d\ota/dz|_0)^2} + \frac{3}{5}+\frac{3}{2}\frac{{\ota}_0/z_{max}}{d\ota/dz|_0}
}\nonumber
\,.\\ 
&& \label{slope}
\end{eqnarray}
The error of the intercept is related to that of the slope through
$\sigma^2_{{\ota}_0} = \sigma^2_{d\ota/dz|_0} \langle z^2\rangle$. 
Calculating the latter average and using Eq.~(\ref{slope}) we obtain
\begin{equation}
\frac{\sigma_{{\ota}_0}}{{\ota}_0 } 
\sim \frac{\sigma_{d\ota/dz|_0}}{ {\ota}_0 }\sqrt{\frac{3}{5}}z_{max}
\,.
\end{equation}

\begin{figure}[tb!]
    \centering
    \includegraphics[width=0.9\columnwidth,clip]{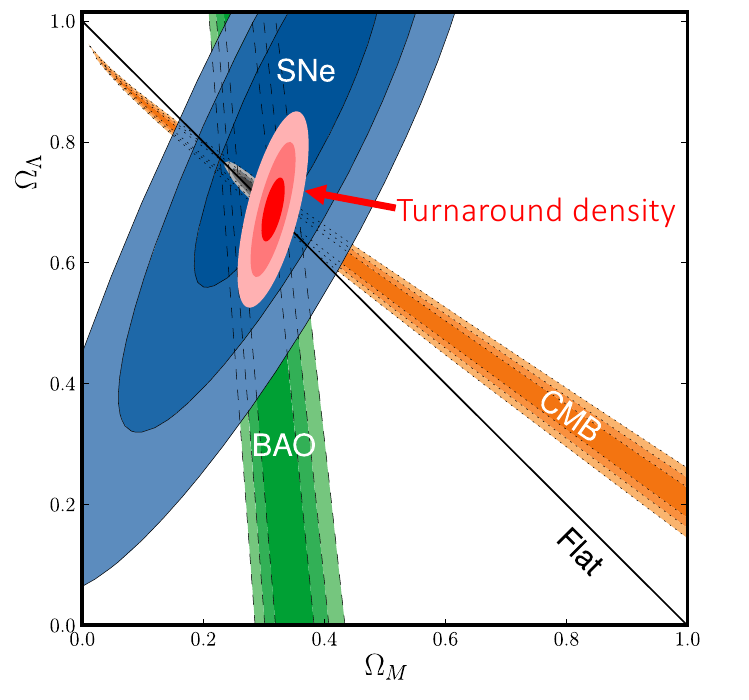}
\caption{Red/pink contours: projected 1, 2, and 3$\sigma$ confidence intervals on $\omo$
  and $\olo$ from turnaround density data only, obtained by combining the constraints from
  Eq.~(\ref{functionf}) and Eq.~(\ref{magick}) shown in 
  Fig.~\ref{constraints}.  As in Fig.~\ref{newline}, the contours are over-plotted on Fig.~10
of \cite{ConstraintsPlot} showing constraints from supernovae
  (blue), the CMB (orange), and baryon acoustic oscillations (green).}\label{contours} 
\end{figure}

Measurements of $\ota$ in
individual clusters will likely have a  relatively poor accuracy ($ f
\sim 0.5$)\footnote{This comes from a typical uncertainty of $\sim 30\%$ in the cluster mass measurement (e.g. \citealp{weak-lensing-mass}), an uncertainty of $\sim 10\%$ in the turnaround radius (comparable to what has been claimed for nearby clusters, e.g., \citealp{LocalGroup,Virgo}), and a $\sim 25\%$ halo-to-halo scatter seen in N-body simulations \citep{Ketal2019}.}.
If concordance $\Lambda$CDM cosmology holds, then ${\ota}_0 = 3.71$
(Eq.~\ref{functionf}) and $d\ota/dz|_0 = - d\ota/da_{ta}|_0 = 5.67$
(Eq.~\ref{magick}). If then 
$\ota$ were measured in the $\sim 42,000$ clusters
with $z_{max}=0.3$ and $M_{200} > 0.6\times 10^{14} {\rm M_\odot}$ in
the 14,000 square degrees of the Sloan Digital Sky Survey (SDSS) III \citep{WHLClusters}
with $f=0.5$,  we could obtain an estimate of ${\ota}_0$
with an uncertainty of $\sim 1.5\%$, and an estimate of $d\ota/dz|_0$
with an uncertainty of $\sim 3.5\%$. These would yield the 
constraints on $\omo$ and $\olo$
shown in Fig. (\ref{newline}) with the red (Eq.~\ref{functionf}) and
purple (Eq.~\ref{magick})
contours. The projected measurement of $\olo$ shown features a $7\%$
accuracy, corresponding to a $14\sigma$ confidence level that $\olo >
0$.  In Fig.~\ref{contours} we have combined the constraints from Eqs.~(\ref{magick}) and 
Eq.~(\ref{functionf}) to obtain confidence levels on the values of
$\omo$ and $\olo$, assuming all uncertainties to be Gaussian and
ignoring non-linear corrections to the low-redshift behavior of
$\ota(z)$. 

\section{Discussion}\label{discussion}
Eq.~\ref{magick} being independent of $\omo$ is a nontrivial
 feature, not shared by any of the currently used probes of the
 cosmological parameters, which makes the evolution of
 $\rho_{ta}$ especially attractive as a direct, local probe of
 $\olo$. The CMB primarily probes the geometry
 of the Universe, so the confidence region on the $\omo-\olo$ plane
 that can be directly derived from it resembles a negative-slope strip
 (Fig.~\ref{newline}, orange), and it is only in combination with
 other probes (e.g., weak lensing, or baryon acoustic oscillations)
 that the CMB reveals a clear preference for $\olo\sim0.7$ (e.g.,
 \citealp{WMAP03, PlanckCosmology2018}). Cluster abundances and baryon
 acoustic oscillations on the other hand are primarily sensitive to
 $\omo$ (as is the present-day value of the turnaround density) and by
 themselves reveal very little about the existence of a cosmological
`` constant (Fig.~\ref{newline}, green and red). This is why
 observational evidence in support of the currently accepted $\omo\sim
 0.3$  predated the general adoption of a non-zero cosmological
 constant  (e.g.~\citealp{oldschool,Neta,shaun}). Finally, type Ia
 supernovae, which, as standard candles, probe the cosmological
 parameters by mapping the redshift dependence of the luminosity
 distance, only constrain the relative values of $\omo$ and $\olo$ if
 they are observed in the low-redshift Universe. This constraint is a
 diagonal positive-slope strip on the $\omo-\olo$ plane
 \citep{perl97}. The slope of this strip changes with increasing
 redshift, and thus only by extending supernovae observations to high
 redshifts ($z\sim1$)  can a measurement of both $\omo$ and $\olo$ be
 obtained \citep{perl95}. By contrast, through Eq.~\ref{magick} we can
 estimate $\olo$ today, with no reference to $\omo$, using the effect
 of $\Lambda$ on galaxy-cluster scales rather than on the Universe as
 a whole, and based on low-redshift observations alone.

 Note that the independence of Eq.~\ref{magick} from $\omo$ holds only
 at $z=0$. Producing an estimate of $\olo$ using Eq.~\ref{magick}
 requires measuring the present-day slope and
 intercept of $\ota(z)$, ideally within a redshift range where
 $\ota$ grows linearly with $z$.
When data from higher redshifts are used, non-linear terms will
introduce some dependence between the $\olo$ and $\omo$  estimates,
which however can be quantified based on Eq.~\ref{functionf}. 
In practice, once measurements of $\ota$ are obtained, a
nonlinear fit will be performed on the predicted $\ota(z)$
parameterized by $\omo$ and $\olo$ (see lower panel of
Fig.~\ref{rho-a}),
which will yield a measurement of
both cosmological parameters (similar to the
contours of Fig.~\ref{contours}).
 For the range $z\leq 0.3$ considered in \S \ref{constraints}, the deviation
of $\ota(z)$ from its linear approximation around
$z=0$ is $<10\%$ for any value of $\olo$: small, but not
subdominant compared to the statistical uncertainties that can
be achieved using the large number of available clusters out to this
distance. For this reason, the predicted shapes of the constraints
from $\ota(z)$ in
Fig.~\ref{newline} should be viewed as approximate. 

The technique discussed here depends on independent measurements of $\ota$ in
different structures, and does not require completeness of the sample
of clusters used. For this reason, the same result can be obtained
with measurements in only a  
fraction of the $z\leq 0.3$ SDSS clusters, if the sample is optimized in
terms of its redshift distribution.
There is also margin for improvement in the
accuracy of the measurement of $\ota$ in individual clusters, for
example by applying quality cuts based on the absence of massive
neighbors \citep{Ketal2019}, mass cuts \citep{weak-lensing-mass}, or
galaxy-number cuts \citep{Virgo, Leesix}. About $500$ well-selected $z<0.3$
clusters, uniformly distributed over redshift, 
with $\ota$ measured with $25\%$ accuracy in each, 
would be enough to 
establish $\olo >0$ at the $5\sigma$ level.  

In estimating the potential accuracy of constraints on $\omo$ and
$\olo$, we assumed all
uncertainties to be statistical. However, before any statement on
cosmology can be made based on measurements of $\ota(z)$, systematic errors have to be
carefully considered as well. For the proof-of-principle calculation
of \S \ref{evolution} we used the model of spherical collapse of a
single structure in an otherwise uniform and isotropic 
background universe. Although \citet{Ketal2019} have shown that the
predictions of spherical collapse for $\ota$ persist in N-body
simulations, they have reported a small systematic shift towards higher
values of $\ota$, due to effects opposing gravity in realistic
cosmic structures (primarily tidal forces from massive neighbors
and, to a much smaller extent, rotation - see also \citealp{BT2019}). 
Whether this shift evolves with redshift and how remains to be
explored with simulations.

Measuring $\rho_{ta}$ in a single cluster requires separate
measurements of the cluster turnaround radius, $R_{ta}$ and of the
cluster mass $M_{ta}$ within $R_{ta}$; then the turnaround density is simply
obtained from $\rho_{ta}= 3 M /4\pi R_{ta}^3$.  Each of $R_{ta}$ and
$M_{ta}$ might also suffer from systematic biases in their
measurement that also have to be quantified and
accounted for, using both mock observations of simulated structures
and cross-calibration
of measurements using different techniques.  The uncertainty
of a single measurement of $\rho_{ta}$ is dominated by that of $R_{ta}$, which in
turn can be obtained from observations of peculiar velocities with
respect to the Hubble flow of cluster member galaxies, provided these
galaxies have distances measured by some indicator other than
redshift. Estimates of the turnaround radius have been attempted in several
nearby structures \citep{LocalGroup,Leesix}, including the Virgo
cluster \citep{Virgo}, and the Fornax-Eridanus complex
\citep{Fornax}. Although 
deriving cosmological parameters from these measurements requires a
careful consideration of uncertainties involved that is beyond of
scope of this work, it is worth noting that these measurements are
consistent with Planck-parameters concordance $\Lambda$CDM.

Deriving $\ota$ from $\rho_{ta}$ in general requires an assumption on the value of
$H_0=h\times 100 {\rm \, km \, s^{-1} Mpc^{-1}}$. However, many
techniques for the 
measurement of cluster masses and cluster radii are themselves calibrated on the
local expansion of the Universe (i.e. they yield masses in
$h^{-1} M_{\odot}$ and radii in $h^{-1} {\rm Mpc}$). Estimates of
$\ota$  obtained from such measurements are independent of $h$.

Extending such analyses to a large number of structures will require the measurement of a large number of cluster masses, redshifts, and
independent distance estimates of member galaxies. However, 
these measurements  are already considered of high cosmological significance and
large-scale campaigns to obtain them are planned or underway. In this context, the turnaround density provides a new promising way to analyze and exploit upcoming large cosmological datasets.  

\begin{acknowledgements}
We thank A. Zezas, K. Tassis,  I. Papadakis, V. Charmandaris,
N. Kylafis, and I. Papamastorakis for valuable discussions, and an anonymous referee
for their careful review and constructive comments. GK acknowledges
support from the European Research Council under the European Union’s
Horizon 2020 research and innovation program, under grant agreement No
771282.  
\end{acknowledgements}

\bibliographystyle{aa}
\bibliography{bibliography}

\end{document}